\documentclass[10pt,english,prd,aps,showpacs,tightenlines,superscriptaddress,amsmath,amssymb,amsfonts]{revtex4-2}
\usepackage[utf8]{inputenc}
\usepackage{babel}
\usepackage{amsmath}
\usepackage{amsthm}
\usepackage{amssymb}
\usepackage{physics}
\usepackage{braket}
\usepackage{slashed}
\usepackage{graphicx}
\usepackage{mathrsfs}
\usepackage{cancel}
\usepackage{amsfonts}
\usepackage{comment}
\usepackage{float}
\usepackage{xcolor}
\begin{document}

\title{Quantum Information Meets High-Energy Physics: Probing Neutrinos and Beyond}

\author{R. Serao}
\email{rserao@unisa.it}
\affiliation{Dipartimento di Fisica ``E.R. Caianiello'' Universit\'a di Salerno,
and INFN -- Gruppo Collegato di Salerno, Via Giovanni Paolo II, 132,
84084 Fisciano (SA), Italy}

\author{G. Torre}
\email{gianpaolo.torre@irb.hr}
\affiliation{Institut Ru\dj er Bo\v{s}kovi\'c, Bijeni\v{c}ka cesta 54, 10000 Zagreb, Croatia}

\author{A. Capolupo}
\email{capolupo@sa.infn.it}
\affiliation{Dipartimento di Fisica ``E.R. Caianiello'' Universit\'a di Salerno,
and INFN -- Gruppo Collegato di Salerno, Via Giovanni Paolo II, 132,
84084 Fisciano (SA), Italy}

\begin{abstract}
This review explores the interplay between quantum information theory and high-energy physics, emphasizing how decoherence effects and unconventional neutrino oscillation patterns may unveil fundamental properties such as the Dirac or Majorana nature of neutrinos and potential CPT violation. It further discusses the use of entanglement measures as novel probes of axion-mediated interactions, outlining interdisciplinary strategies to test the limits of the Standard Model and explore new physics beyond it.

\end{abstract}
\maketitle

\section{Introduction}

In high-energy physics, the behavior and dynamics of elementary particles are most commonly described by scattering amplitudes between asymptotic states in the framework of flat spacetime. This approach, known as the S-matrix formalism \cite{Smat}, has proven to be highly effective, particularly in the context of collider physics, where it provides a clear and powerful tool for analyzing particle interactions. The success of this framework, however, is not without its limitations. When applied to situations where spacetime curvature, long-range correlations, or environmental influences play a significant role, the S-matrix approach encounters conceptual challenges. In such cases more sophisticated approaches are required to capture the full complexity of particle dynamics in these non-trivial settings. At the same time, while the Standard Model of particle physics offers a robust framework for understanding a broad range of phenomena involving fundamental particles and their interactions, it is important to acknowledge that it does not constitute the final theory of elementary particles. Indeed, although the Standard Model has been remarkably successful in explaining many observed phenomena, there remain unresolved questions and unexplained phenomena that lie outside its purview. For example, phenomena such as particle mixing \cite{neu1,neu2,Pontecorvo,Pontecorvo1,neu5,neu6,neu7,BSM1,BSM2,BSM3,BSM4,BSM5,BSM6,BSM7,BSM8}, which involves the transformation of one type of particle into another, and the quantum features of gravitation \cite{QG, QG1, QG01, QG02, QG2, QG3, QG4, QG5, Penrose} are not fully accounted for within the Standard Model. It is therefore necessary, on the one hand, to advance our knowledge of particle physics by abandoning certain restrictive assumptions and models, and on the other to adopt new analytical tools.

Among the various approximations, perhaps the most pivotal to relax is the idealized assumption of an isolated system, neglecting interactions with its surrounding environment. Indeed, this inevitable interaction leads to a range of non-trivial effects, such as dissipation and decoherence, that cannot be neglected. The theory of Open Quantum Systems (OQS)~\cite{OQS1,OQS3,OQS4} has gained significant attention over recent decades, as it offers powerful tools for addressing the complexities introduced by environmental coupling thereby enabling more accurate predictions of quantum system behavior in real-world settings, such as experimental setups. Originally developed within the framework of quantum optics, it has been used to investigate the role of dissipation and irreversibility in various particle physics phenomena \cite{Marco1,Marco2,Marco3,Marco4}, with the initial motivation being the analysis of quantum gravity effects. Indeed, spacetime, treated as a dynamical variable, undergoes fluctuations of both classical and quantum in nature, and these fluctuations lead to decoherence in quantum systems (see~\cite{Bassi2017} for a review). Moreover, incorporating gravity into the quantum systems description is not merely a step toward more realistic modeling; it represents a low-energy quantum mechanical approach to reconciling quantum mechanics with general relativity, potentially circumventing the need for more sophisticated theories, such as string theory~\cite{Zwiebach2009} or loop quantum gravity~\cite{Gambini2011}. Finally, beyond gravitational effects, modeling quantum systems as open systems is crucial for uncovering novel aspect of physical phenomena~\cite{Marco1,Marco2,Marco3,Marco4}.

The environment-induced decoherence mechanism and the corresponding quantum to classical transition cannot be explained without resorting to the concept of entanglement~\cite{Zurek2007, Nielsen2010}. Indeed, even the basic measurement process requires the system and the measurement apparatus to correlate, causing system wave function to collapse. The environment can be understood as the collection of degrees of freedom that are dynamically coupled to the system of interest and become entangled with it. It is this continuous interaction, effectively a form of monitoring, that drives the system towards decoherence~\cite{Zurek2007}. Entanglement thus constitutes a fundamental element in addressing the dynamics of open quantum systems and has sparked growing interest within the particle physics community. Moreover, from a broader perspective, the application of quantum information concepts and techniques is becoming increasingly pervasive in physics, ranging from many-body systems~\cite{Amico2008}, to black hole physics~\cite{Solodukhin2011}, and the AdS/CFT correspondence~\cite{Hubeny2015}, emerging as a new paradigmatic framework for exploring and understanding physical phenomena, and representing a fundamental ingredient in the future research road-map of particle physics (see~\cite{Pei2025} for a review). This review aims to highlight the role of entanglement as a fundamental tool for investigating physics beyond the Standard Model in the more realistic scenario of a non-isolated system.

We start considering particle mixing, that occurs when particles identified by definite flavor states are, in fact, quantum superpositions of mass eigenstates with slightly different masses~\cite{Buras,PDG}. This small mass difference makes the system highly sensitive to external perturbations, such that even weak interactions can induce observable modifications in the oscillation frequencies. This enhanced sensitivity forms the foundation of several experimental proposals designed to probe the physical consequences of flavor mixing \cite{PDG}. As paradigmatic example of particle mixing we consider neutrinos mixing, since they interact only via gravity and the weak force. Moreover, although the weak force is stronger than gravity, its range is extremely limited (on the order of
$d=10^{-16}-10^{-18}$ meters), meaning it can be neglected over distances larger than $d$, where gravitational interactions dominate. In this context, the quantum information approach provides a promising avenue for advancing our understanding of neutrinos.
Indeed, although the key neutrinoless double beta decay could provide insight into the nature of neutrinos and several experimental setups has been proposed to detect this process~\cite{Gomez, Dolinski, Elliott}, no conclusive results have yet been obtained.
The inclusion of decoherence effects in neutrino oscillation formulas can introduce dependencies on the Majorana phase, a characteristic that could be used to distinguish between Dirac and Majorana neutrinos (see Section~\ref{Neutrinos} and~\cite{Marco1, Marco2, Marco3, Marco4} for an in deepth discussion). Furthermore, this approach offers a pathway to exploring how off-diagonal dissipators can be used to identify the nature of neutrinos, with one important consequence being the potential violation of CPT symmetry~\cite{CPT1,CPT2,CPT3,CPT4,CPT5,CPT6}.

The second paradigmatic case we  consider concerns the search for axions, hypothetical light ($m_a\simeq 10^{-6}-10^{-2}$)~\cite{AX13, AX14}, neutral particles that interact extremely weakly with ordinary matter, and is intimately related to the longstanding and unresolved strong CP problem~\cite{AX2, AX3, AX4, AX5, AX6, AX7, AX8, AX9}.
Indeed, in the Standard Model of particle physics, the QCD Lagrangian contains a term that could, in principle, break the combined symmetry of charge conjugation (C) and parity (P) whereas no such violation is tested in QCD processes ( see Ref.~\cite{PDG} and references therein). A possible solution was proposed by Peccei and Quinn~\cite{AX2, AX3}, who introduced a new global symmetry that is spontaneously broken at a high energy scale. This mechanism leads to the emergence of a new pseudo-Nambu--Goldstone boson, subsequently identified as the axion~\cite{AX2,AX3,AX4, AX5, AX10, AX11, AX12,AX13,AX14}. Axions have also emerged as compelling dark matter candidates~\cite{AX15, AX16, AX17, AX18, AX19}. If they exist, axions produced in the early universe, e,g, via the decay of the Peccei-Quinn symmetry, could constitute a significant fraction of cold dark matter. Despite their extremely low mass and weak coupling to ordinary matter, they could exert a gravitational influence on the large-scale structure of the universe, and its existence might be inferred through indirect observations. Owing to these strong theoretical motivations, numerous experiments have been devised to search for axion-like particles (ALPs), despite the extreme weakness of their interactions. Most efforts rely on axion–photon coupling~\cite{AX20,AX21,AX22,AX23,AX24,AX25}. Moreover, more recent approaches explore quantum geometric phases and axion–photon mixing~\cite{AX26, AX27} and from studying axion–nucleon and axion–lepton interactions \cite{AX28,AX29}. Nonetheless, no experimental evidence for ALPs has been found so far. In this review we show that it is possible to detect indirectly the axion based on the entanglement generated between two fermions through axion-mediated interactions, namely the entanglement could serve as indirect evidence of pseudoscalar interactions, thereby supporting the existence of axions and ALPs.

The paper is structured as follows. In Section~\ref{OQSAndEnt}, we recall the fundamental concepts and formalism used to describe open quantum systems, namely, systems that are not isolated but interact with their surrounding environment. In particular, we introduce the Gorini–Kossakowski–Lindblad–Sudarshan (GKLS) master equation, which is widely employed under the assumptions of weak coupling and Markovian dynamics. These tools are then applied in subsequent sections to the study of neutrino oscillations in the presence of gravitational interactions and, more generally, under decoherence effects. Finally, we outline the key concepts and methods related to entanglement, which underlies the emergence of decoherence and the associated phenomenology. In Section~\ref{Neutrinos}, we investigate two-flavor neutrino oscillations. We begin by reviewing the idealized case of a perfectly isolated system, and then demonstrate that, when mutual gravitational interactions between neutrinos are taken into account, translational symmetry is broken, which in turn leads to CPT symmetry violation. We then consider a more general scenario incorporating decoherence effects, described by the Gorini--Kossakowski--Lindblad--Sudarshan (GKLS) master equation. Within this framework, we identify conditions under which CP symmetry, and consequently CPT symmetry, is violated. This phenomenon is further connected to the possibility of probing the fundamental nature of the neutrino, specifically whether it is a Dirac or a Majorana particle. In Section~\ref{axions}, we demonstrate how the emergence of entanglement at times $t > 0$ between two fermions, initially prepared in a product state at $t = 0$, can be linked to a hypothetical interaction mediated by an axion field. Finally, in In Section~\ref{Conclusions}, we summarize our main findings and outline possible directions for future research, highlighting how our results may contribute to ongoing efforts in theoretical modeling and experimental investigations in particle physics through the lens of entanglement.

\section{Open quantum systems and entanglement}\label{OQSAndEnt}

In this Section, we briefly introduce the fields of open quantum systems and entanglement theory, with an emphasis on aspects relevant to the subsequent discussion. For a more comprehensive treatment, we refer the reader to the literature~\cite{10.1093/acprof:oso/9780199213900.001.0001,Nielsen2010,OQS1}.

\begin{figure}[t!]
    \centering
\includegraphics[width=0.4\textwidth]{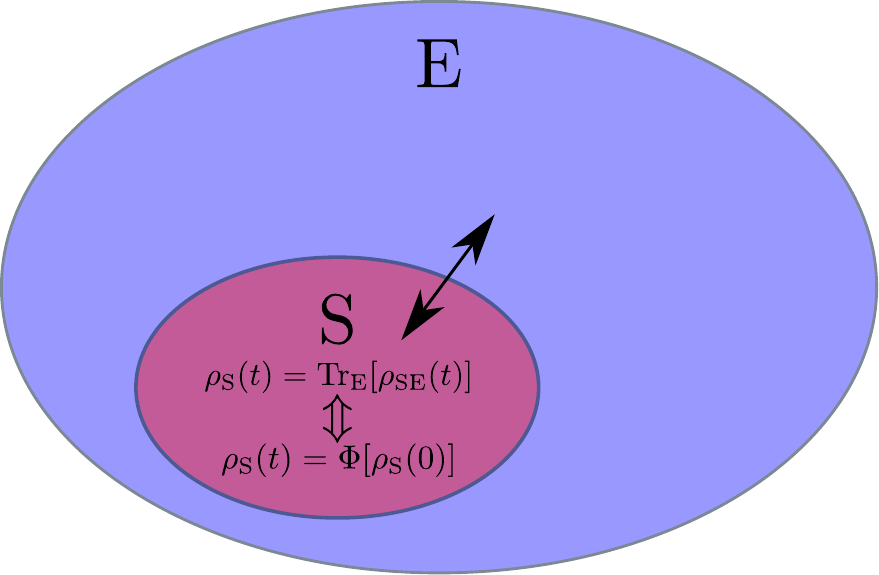}
    \caption{A realistic description of the dynamics of a quantum system $\textrm{S}$ must incorporate its interaction with the surrounding environment $\textrm{E}$. While the joint system–environment evolution, governed by the Hamiltonian $\mathcal{H}_{\textrm{SE}} = \mathcal{H}_{\textrm{S}} + \mathcal{H}_{\textrm{E}} + \mathcal{H}_{\textrm{Int}}$ (with $\mathcal{H}_{\textrm{Int}}$ denoting the interaction term), is unitary, the reduced dynamics of the system is non-unitary as a consequence of this interaction. The reduced density matrix describing the system at time $t > 0$ can be obtained by tracing out the environmental degrees of freedom from the evolved total state, or equivalently, by applying a dynamical map to the initial system state. This map must be completely positive (CP) in order to preserve the physicality of the state~\cite{10.1093/acprof:oso/9780199213900.001.0001}.
 } 
    \label{fig:SEScheme}
\end{figure}

A realistic description of the evolution of a quantum system $\textrm{S}$ must necessarily account for its interaction with the surrounding environment \textrm{E}, with which it exchanges energy and/or information. As a result, the evolution is no longer unitary, and the reduced state of the system cannot be described by a pure state, but rather by a mixed state represented by a density matrix. Conceptually, the state of the system at a given time $t > 0$ can be obtained by considering the global unitary evolution of the combined system–environment ($\textrm{SE}$) and subsequently tracing out the environmental degrees of freedom ($\rho_{\textrm{S}}(t)=\textrm{Tr}_\textrm{E}[\rho_{\textrm{SE}}(t)]$)(see Fig.\ref{fig:SEScheme}). This is the most straightforward approach, and it is indeed the method we adopt in Section~\ref{Neutrinos} to obtain the reduced density matrix describing the evolution of a single neutrino interacting with a surrounding neutrino environment. Nevertheless, a dedicated mathematical framework has been developed to account for such situations. This mathematical description of the interaction is provided by a dynamical map, which evolves the initial state of the system according to 
$\rho_{\textrm{S}}(t) = \Phi[\rho(0)]$. In order to preserve the positive semi-definiteness of the density matrix for all possible initial states, including those entangled with ancillary systems, the map $\Phi$ must be completely positive (CP)~\cite{Nielsen2010}. This requirement ensures that the physicality of the quantum state is maintained throughout the evolution.

Although the dynamical map formalism is the most general framework, capable of describing both Markovian and non-Markovian processes, as well as cases involving strong system–environment coupling, in many practical situations, the weak-coupling and memoryless Markovian approximations are sufficient. Under these conditions, the reduced dynamics of the system admits a simple microscopic description in terms of the Gorini–Kossakowski–Lindblad–Sudarshan (GKLS) master equation~\cite{10.1093/acprof:oso/9780199213900.001.0001}
\begin{equation}\label{GKLDEq}
    \dfrac{d\rho_S(t)}{dt}=-i[\mathcal{H},\rho_S(t)]+ \mathcal{L}_{\textrm{GKLS}}[\rho_S(t)]
\end{equation}
where $\mathcal{H}$ is the full system plus environment Hamiltonian, and $\mathcal{L}_\textrm{GKLS}$ is the dissipator
\begin{equation}
    \mathcal{L}_\textrm{GKLS}[\rho_S(t)]= -i [\mathcal{H},\rho_S(t)]+\sum_{i,j=1}^{N^2-1} \gamma_{ij}\left[L_i \rho_S(t) L_j - \dfrac{1}{2}\left\{L_j L_i, \rho_S(t)\right\}\right].
\end{equation}
%\begin{equation}
%    \mathcal{L}_\textrm{GKLS}= -i [\mathcal{H},\rho_s(t)]+\sum_{\alpha\beta} \gamma_{\alpha,\beta}\left[L_\alpha \rho_s(t) L_\beta - \dfrac{1}{2}\left\{L_\beta L_\alpha, \rho_s(t)\right\}\right].
%\end{equation}
The jump operators $L_i$, with $i = 1, \ldots, N$, where $N$ is the dimension of the system’s Hilbert space ($\dim \mathcal{H}_{\textrm{S}} = N$), characterize the dissipative part of the dynamics. The eigenvalues of the positive semidefinite matrix $\gamma$ represent the relaxation rates associated with the various decay modes of the open quantum system. In Section~\ref{Conclusions}, we make use of Eq.~\eqref{GKLDEq} to describe the non-unitary evolution of two-flavor neutrinos.

The other quantum mechanical concept we shall rely on is entanglement~\cite{RevModPhys.81.865, Nielsen2010}. Although the existence of entangled states has been known since the early days of quantum mechanics~\cite{PhysRev.47.777,Schrödinger_1935}, it took several decades before the phenomenon was systematically studied through the formulation of Bell inequalities~\cite{PhysicsPhysiqueFizika.1.195}, and subsequently confirmed by experimental tests, most notably those performed by Aspect and collaborators~\cite{PhysRevLett.49.1804}. A major boost in the study of entanglement came in the 1990s, with the realization that it constitutes a fundamental resource for quantum technologies such as quantum teleportation~\cite{PhysRevLett.70.1895}, superdense coding~\cite{PhysRevLett.69.2881}, and quantum cryptography~\cite{BENNETT20147,PhysRevLett.67.661}. Today, entanglement is widely regarded as one of the pillars of what is often referred to as the ``Second Quantum Revolution''~\cite{DowlingMilburn2003}. The definition and quantification of entanglement are most naturally introduced in the simplest case of a bipartite system. Moreover, since we are dealing with a system--environment setting, we will illustrate the basic concepts in this framework.

According to the postulates of quantum mechanics, the Hilbert space of the total system, comprising both the system and its environment, is given by the tensor product of the individual Hilbert spaces, $H = H_{\textrm{S}} \otimes H_{\textrm{E}}$. As a consequence of the superposition principle, the state of the composite system can be expressed as $\ket{\psi_{\textrm{SE}}} = \sum_{\boldsymbol{i}_n} c_{\boldsymbol{i}_n} \ket{\boldsymbol{i}_n}$, where $\boldsymbol{i}_n = (i_1, i_2)$ is a multi-index labeling the basis states of the two subsystems. The indices $i_1$ and $i_2$ run over orthonormal bases of $H_{\textrm{S}}$ and $H_{\textrm{E}}$, respectively. In general $\ket{\psi_{\textrm{SE}}}\neq \ket{\psi_{\textrm{S}}}\otimes \ket{\psi_{\textrm{E}}}$, namely, at variance with the classical case, the total state is not always a product state of the $2$ separate systems. This general impossibility captures the phenomenon of entanglement, which, in contrast to the classical case, allows for the creation of exponentially large superpositions with only linear physical resources.

In our setting, the initial system--environment state is assumed to be pure and factorized, $\ket{\psi_{\textrm{SE}}} = \ket{\psi_{\textrm{S}}} \otimes \ket{\psi_{\textrm{E}}}$. If no interaction occurs between the system and the environment, the state of the system remains pure even after discarding the environmental degrees of freedom. It is then the interaction with the environment that generates entanglement. As a result, tracing out the environment leads to a mixed reduced state $\rho_\textrm{S}$. We can then measure the entaglement content of $\rho_\textrm{S}$ through its purity
\begin{equation}\label{purity}
    \mathcal{P}(\rho_{\textrm{S}}(t)) = \textrm{Tr}[\rho_{\textrm{S}}^2(t)].
\end{equation}

We apply this expression in Section~\ref{Neutrinos} to quantify the entanglement induced by gravitational interactions between a neutrino and its surrounding neutrino background. Furthermore, in Section~\ref{axions}, we consider the Rényi-2 entropy, defined as
\begin{equation}\label{renyi}
    S_2(\rho_{\textrm{S}}(t)) = -\ln[\mathcal{P}(\rho_{\textrm{S}}(t))],
\end{equation}
to evaluate the entanglement between two fermions as an indirect signature of their interaction, mediated by an axion.

\section{Neutrino oscillation, CPT symmetry violation, and the nature of the neutrino}\label{Neutrinos}

Since we are interested in neutrino mixing, in this Section we briefly review some aspects of neutrinos oscillations relevant to this case, referring the reader to the literature~\cite{PDG} for further details. Furthermore, we consider only the oscillation between two neutrinos with well-defined flavor, and refer the reader to the literature for the general three-flavor oscillation framework~\cite{Pontecorvo, Pontecorvo1}.
The significance of this phenomenon lies in the fact that the observation of neutrino oscillations provides experimental evidence that neutrinos have mass, a feature not accounted for within the Standard Model.

In the most general case, the mixing relations can be written as~\cite{Pontecorvo, Pontecorvo1} 
\begin{equation}\label{neutrinoMix}
\begin{split}
    &\ket{\nu_e}=\cos[\Theta] \ket{\nu_1}+ e^{i\phi} \sin[\Theta] \ket{\nu_2}\\
    & \ket{\nu_\mu}=-e^{-i\phi}\sin[\Theta] \ket{\nu_1}+ \cos[\Theta] \ket{\nu_2}
\end{split}
\end{equation}
Where $\Theta$ is the mixing angle, $\phi$ is the Majorana phase, and $\ket{\nu_1}, \ket{\nu_2}$ are the states with definite mass $m_1, m_2$ respectively.
Eq.s~\eqref{neutrinoMix} are compatible with two possible natures of the neutrino, namely, it may be either a Dirac or a Majorana particle. In the Dirac scenario ($\phi=0$), neutrinos and antineutrinos are distinct particles, as is the case for charged fermions. In contrast, in the Majorana framework, neutrinos are their own antiparticles, implying lepton number violation and offering a natural explanation for the smallness of neutrino masses through the seesaw mechanism~\cite{seesaw}.
If we now consider the evolution of the neutrino as a closed systems, namely neglecting any interaction with the surrounding environment, the Hamiltonian describing the mixed fields can be written as
\begin{equation}\label{NIH}
    H_{\textrm{mix}}=E+\frac{c^2}{2E}(m_1^2\ket{\nu_1}\bra{\nu_1}+m_2^2 \ket{\nu_2}\bra{\nu_2})=\omega_0\sigma^z,
    \end{equation}
where $\omega_0=\frac{c^2}{2E}(m_1^2-m_2^2)$, $E$ is the energy of the particle, which is assumed to be $m_i\ll E$, and where in the final step we have neglected terms proportional to the identity. Under the effect of the Hamiltonian Eq.~\eqref{NIH} the time evolution of a generic flavor state $\ket{\nu_A}$ (with $A=e,\mu$) leads to a non vanishing probability $P_{\nu_A\rightarrow \nu_B }$ with $A\neq B$ ($B=e,\mu$), to observe a change in the flavor state. It is straightforward to show~\cite{Pontecorvo, Pontecorvo1} that the probability to have a flavor oscillation is given by 
\begin{equation}\label{OscillationProb}
    P_{\substack{\nu_A\rightarrow\nu_B \\ \nu_B\rightarrow\nu_A}}=|\bra{\nu_{A/B}}e^{-i\mathcal{H}_{\textrm{mix}}t}\ket{\nu_{B/A
}}|^2=\sin^2[2\Theta]\sin[\omega_0t].
\end{equation}
 This results is the well-known Pontecorvo formula. From Eq.~\eqref{OscillationProb} we see that it in not possible to distinguish between the Dirac and Majorana nature of the neutrinos, as it does not depend on the Majorana angle $\phi$. Furthermore, these oscillations do not violate CPT symmetry, since CPT implies that the transition probability of neutrinos and antineutrino are equal under particle-antiparticle exchange, combined with space and time inversion.

We now examine the scenario in which a neutrino undergoes gravitational interactions with a surrounding neutrino background. In particular, we assume that the particles travel in the space with the same energy and in the same direction, and the validity of the equivalence principle between inertial and gravitational mass.  The problem can be addressed in full generality by considering the case of \( N \) neutrinos interacting gravitationally (see~\cite{Simonov2019} for a comprehensive treatment). In this review, we focus on the simplest case of two neutrinos, which is sufficient to capture the essential physics, and briefly comment on the extension to more general scenarios.

The Hamiltonian governing the evolution is given by
\begin{equation}\label{twoHam}
    \mathcal{H}=\omega (\sigma_1^z+\sigma_2^z)+\Omega\; \sigma_1^z\cdot\sigma_2^z,
\end{equation}
where $\omega= \omega_0+ g(m_1^2-m_2^2), \Omega=g(m_1-m_2)^2$, $g=-\frac{G}{4d_{i,j}}$, $G$ being the gravitational constant, and $d$ is the distance between the two particles. As discussed in Section~\ref{OQSAndEnt} at $t>0$ the evolution of the $1$($2$)-th neutrino will not be unitary, due to the reciprocal interaction in Eq.~\eqref{twoHam}. To obtain the expression for its reduced density matrix, we then evolve the full system and then trace out the degrees of freedom associated with the $2$($1$)-th neutrino. It is straightforward to show that the purity Eq.~\eqref{purity} reads~\cite{Simonov2019}
\begin{equation}\label{twoPurity}
    \mathcal{P}(\rho_{1(2)}(t))=1-\dfrac{1}{2}\sin^4(2\theta)\sin^2(2t\Omega)
\end{equation}
Fom Eq.~\eqref{twoPurity} we see that in presence of mixing ($\theta \neq 0,\pi/2$, and $t\neq (n\pi)/2\Omega$) the effect of the garavitational interaction is to entangle the two particle. reducing the purity. This fact may be regarded as supporting the quantum character of gravity, given that quantum correlations and entanglement cannot arise through classical interactions, but only via a quantum channel~\cite{PhysRevLett.83.3081}.

The presence of entanglement also leads to a violation of time-reversal T and then CPT symmetry. In fact, let us suppose to have two identical systems, except for the initial state, namely $\ket{\psi_A(0)}=\ketbra{\nu_A}{\nu_A}, \ket{\psi_B(0)}=\ketbra{\nu_B}{\nu_B}$, with $A\neq B$. If we now compute the probability for these two cases we obtain~\cite{Simonov2019}
\begin{equation}\label{twoProbability}P_{\substack{\nu_A\rightarrow\nu_B \\ \nu_B\rightarrow\nu_A}} = \dfrac{1}{2}\sin^2(2\theta)[\cos(2\omega t)\cos(2\Omega t)\pm\cos(2\theta)\sin(2\omega t)\sin(2\Omega t)].
\end{equation}
From Eq.~\eqref{twoProbability} we see that the CP symmetry is preserved, since these expressions do not depend on the Majorana phase $\phi$. On the contrary, we see that $\Delta_\textrm{T}=P_{\nu_A\leftrightarrow\nu_B} - P_{\nu_B\leftrightarrow\nu_A}\neq 0$ (it is zero only for $t\neq \pi/\Omega$ and $k\pi/4, k\in\mathbb{Z}$), namely the time reversal symmetry is violated and consequently the CPT one. We have thus demonstrated that gravity, in a system of self-interacting mixed particles, leads to a violation of CPT symmetry. This violation is linked to the generation of entanglement among the constituents of the system, which arises due to a mass difference between the free fields. Furthermore, owing to the additive nature of the gravitational interaction, the degree of CPT violation scales proportionally with both the number of particles in the system and its density.

An alternative approach, discussed in the literature in the context of decoherence effects in neutrino mixing, involves describing the process using the GKLS equation (Eq.~\eqref{GKLDEq}), which is applicable to long-baseline experiments. In fact, considering the mixing between two flavours, the dynamics on one of the neutrinos can be analyzed by studying the mathematical properties of the dissipator, without specifying any particular form of the Hamiltonian~\cite{Marco4}. Expanding Eq.~\eqref{GKLDEq} in the $SU(2)$ basis we get~\cite{Marco4}
\begin{equation} \label{Lin}
    \frac{d \rho_\lambda}{dt}\sigma_\lambda=2\epsilon_{ijk}H_i\rho_j(t)\sigma_\lambda\delta_{\lambda k}+D_{\lambda\mu}\rho_\mu(t)\sigma_\lambda
\end{equation}
with $\rho_\mu=Tr(\rho \sigma_\mu),\mu\in[0,3]$ and $D_{\lambda\mu}$ is a $4\times 4$ matrix.
The most general form for $D_{\lambda\nu}$ is obtained imposing that the first raw and the first column of $D_{\lambda\mu}$ is equal to 0 to ensure probability conservation, obtaining
\begin{equation}\label{Diss}
    D_{ij}=-
    \begin{pmatrix}
        \gamma_1 &\alpha&\beta\\
        \alpha &\gamma_2 &\delta\\
        \beta&\delta&\gamma_3
    \end{pmatrix},
\end{equation}
with $\alpha,\beta \in \mathbb
R$ real, and $\gamma_i >0, \textrm{for } i,j = 1, \ldots, 3$. From Eq.~\eqref{Diss} it is clear that different evolutions can be considered according to its specific form (see~\cite{Marco3} for a complete classification). In particular we consider the simplest case of non-diagonal form of Eq.~\eqref{Diss}, putting $\beta=\delta=0$, $\gamma_1=\gamma_2=\gamma$, and $\vert \alpha \vert\leq \gamma_3/2 \leq \gamma \; \forall t$, due to the complete positivity condition. By solving Eq.~\eqref{Lin} through Eq.~\eqref{Diss} it is possible to show that~\cite{Marco4}
\begin{equation}\label{openMixing}
    \rho(t) = \frac{1}{2}
    \begin{pmatrix}
        \rho_0(t) + \rho_3(t) & \rho_1(t) - i\rho_2(t) \\
        \rho_1(t) + i\rho_2(t) & \rho_0(t) - \rho_3(t)
    \end{pmatrix}, \qquad 
    \begin{aligned}
        \rho_1(t) &= e^{-\gamma t} 
        [
            \rho_1(0) \cosh(\Omega_\alpha t)
            - \rho_2(0) \frac{\sinh(\Omega_\alpha t)}{\Omega_\alpha} \Gamma_-
        ], \\
        \rho_2(t) &= e^{-\gamma t} 
        [
            \rho_1(0) \frac{\sinh(\Omega_\alpha t)}{\Omega_\alpha} \Gamma_+
            + \rho_2(0) \cosh(\Omega_\alpha t)
        ], \\
        \rho_3(t) &= \rho_3(0) e^{-\gamma_3 t}.
    \end{aligned}
\end{equation}
and where $\Gamma_\pm =\alpha+\Delta M^2/2E, \Omega_\alpha=\sqrt{\alpha^2-\Delta M^4/4E^2}$. The components $ \rho_i(0) $, with $ i = 1, \ldots, 3 $, depend on the initial density matrix of the neutrino in a definite flavor state. From Eqs.~\eqref{openMixing} and~\eqref{neutrinoMix}, one can compute the probability of flavor oscillation, $ P_{\nu_A \rightarrow \nu_B} = \textrm{Tr}[\rho_A(0)\rho_B(t)] $, for $ A \neq B $, as well as the corresponding process for antineutrinos, $ P_{\bar{\nu}_A \rightarrow \bar{\nu}_B}=\textrm{Tr}[\rho_{\bar{A}}(0)\rho_{\bar{B}}(t)] $, and from these the violation of the CP symmetry
\begin{equation}
    \Delta_{\textrm{CP}}^\textrm{M} = P_{\nu_A \rightarrow \nu_B} -  P_{\bar{\nu}_A \rightarrow \bar{\nu}_B} = -\sin^2(2\theta)\dfrac{\alpha\sin(2\phi\sin(\Omega_\alpha t))}{\Omega_\alpha} e^{-\gamma t},
\end{equation}
where $\Omega_\alpha=\sqrt{\alpha^2-\Delta^2}$, with $\Delta=\Delta m^2 / (2 E)$. Moreover, for the case of transitions preserving their flavors we have $\Delta_{\textrm{CP}}^\textrm{A$\leftrightarrow$A} = P_{\nu_A \rightarrow \nu_A} -  P_{\bar{\nu}_A \rightarrow \bar{\nu}_A} = - \Delta_{\textrm{CP}}^\textrm{M}$. Hence, the observed CP violation arises solely from decoherence effects and pertains exclusively to Majorana neutrinos ($\phi\neq 0$). In contrast, Dirac neutrinos do not exhibit any CP violation in the survival probabilities. These results clearly reveal an asymmetry between neutrinos and antineutrinos, which stems from the presence of the Majorana phase: specifically, when the Majorana phase is set to zero, this asymmetry vanishes. This behavior implies that if neutrinos are indeed Majorana particles, then their oscillations would exhibit CP violation, whereas this would not occur if they were Dirac particles.

\section{Probing axion by means of entanglement }\label{axions}

In this Section, we examine the hypothetical interaction between two fermions mediated by an axion field, with the goal of establishing a method to infer the existence of the axion through its entanglement-generating effects.
From a quantum information perspective, it is well known that any nontrivial quantum channel—that is, any physical interaction capable of transmitting information—necessarily gives rise to quantum correlations in the form of entanglement \cite{QC1,QC2}. Accordingly, if a bipartite fermionic system is prepared in an initial product state and a subsequent measurement reveals the presence of entanglement between the two fermions, this implies that an interaction must have taken place during the evolution. However, it must be emphasized that fermions inherently experience various other fundamental interactions, such as electromagnetic or weak forces, which can likewise induce entanglement over time. To isolate the contribution due specifically to the axion-mediated interaction, it is therefore essential to employ suitable shielding techniques, dynamical decoupling protocols, or other experimental strategies that suppress or account for entanglement generation from standard interactions \cite{QC3}. By carefully selecting appropriate time windows and parameter regimes, one can ensure that the residual entanglement due to known interactions is either negligibly small or effectively vanishing, thereby allowing any nonzero entanglement to be ascribed, within experimental uncertainty, to the hypothesized axion exchange.

The two body potential due to axion exchange is:
\begin{equation}
    V(\mathbf{r})=-\frac{g_p^2 e^{-mr}}{16\pi M^2}\biggl[\mathbf{\sigma_1}\cdot\mathbf{\sigma_2}\biggl(\frac{m}{r^2}+\frac{1}{r^3}+\frac{4}{3}\pi\delta^3(\mathbf{r})\biggr)-(\mathbf{\sigma_1}\cdot\mathbf{\hat{r}})(\mathbf{\sigma_2}\cdot\mathbf{\hat{r}})\biggl(\frac{m^2}{r}+\frac{3m}{r^2}+\frac{3}{r^3}\biggr)\biggr]
\end{equation}
where $g_p$ is the effective axion-fermion coupling constant, $m$ is the axion mass, $r$ is the modulus of the relative distance between the fermions, $\mathbf{\sigma_i}$ is the three dimensional vector of Pauli operators defined on the $i$-th fermion and $M$ is the fermion mass. 
This potential leads to the following interaction hamiltonian:
\begin{equation}
    H_p=-\frac{g^2_p e^{-mr}}{16\pi M^2r^3}[m^2r^2\sigma_1^z\sigma_2^z+(mr+1)\Lambda]
\end{equation}
where $\Lambda=2 \sigma^z_1\sigma^z_2-\sigma^x_1\sigma^x_2-\sigma^y_1\sigma^y_2$.
In order to quantify the entanglement, we deal with a system made of two identical spin-$\frac{1}{2}$ fermions and study the time evolution.
We write the state of the system as 
\begin{equation}
    \Psi(\mathbf{r_1},\mathbf{r_2},s_1,s_2;t)= R(\mathbf{R},\mathbf{r};t)\psi(s_1,s_2;t)
\end{equation}
where $R$ is a function which depends on the center of mass position $\mathbf{R}$ and $\mathbf{r}=\mathbf{r_1}-\mathbf{r_2}$ is the relative position of the two fermions.
The spin wave function is the product space $H_1^{spin}\bigotimes H_2^{spin}$. 
In some specific condition we can neglect the spatial wave function and focus on the spin part \cite{QC3}.
Initially the spin state is fully separable:
\begin{equation}
    \begin{split}
        \ket{\psi(0)}&=\ket{\phi(0)}_1\otimes \ket{\phi(0)}_2\\
        \ket{\phi(0)}_i&=\cos(\theta)\ket{\uparrow}_i+e^{i\phi}\sin(\theta)\ket{\downarrow}_i
    \end{split}
\end{equation}

where we have swapped to a more clear Dirac notation and we have dropped the spin arguments $s_1,s_2$.
 To detect entanglement caused specifically by axions between two fermions, other sources of entanglement must be minimized. Strong and weak nuclear forces can be ignored at sufficiently large distances ($r>10^{-12}$ m). Gravitational and electrostatic interactions only produce global phase shifts and do not create entanglement since they do not depend on spin. However, the magnetic dipole–dipole interaction remains a relevant source that needs to be addressed, whose Hamiltonian is
\begin{equation}
    H_\mu=-\frac{1}{4 \pi r^3}\frac{g^2q_e^2}{16M^2}\Lambda,
\end{equation}
with g is the g-factor and $q_e$ the charge of the electron.
This dipole–dipole magnetic interaction, mediated by massless photons, is long-range and sensitive to spin states, meaning it can contribute to entanglement. However, by carefully choosing the time interval for the measurement, its effect can be minimized.

The system evolves under a unitary operator $\ket{\psi(t)}=U(t)\ket{\psi(0)}$ determined by the total Hamiltonian, which includes both the magnetic and axion contributions:
\begin{equation}
\begin{split}
    U(t)&=e^{-itH_T}\\
    H_t&=-\frac{A}{r^3}[\Lambda+Be^{-mr}(m^2r^2\sigma_1^z\sigma_2^z+\Lambda(mr+1)]
\end{split}
\end{equation}
where $A=\frac{g^2q_e^2}{64 \pi M^2}$ is the strenght of the magnetic interaction and $B=\frac{4g_p}{\alpha g^2}$ is the strenght of the axion interaction.
To quantify the entanglement between the two fermions we consider the Rényi-2 entropy, since it has the advantage to be experimentally accessible.

From Eq.~\eqref{renyi} we have
%related to the experimentally accessible quantity Rényi-2 entropy Eq.~\eqref{renyi},\textcolor{red}{!!!!!!!!!!!!!!!!!!!1} defined as $S_2=-\ln(\mathcal{P}(\rho_i(t)))$ where $\rho_i(t)$ is the reduced density matrix and $\mathcal{P}(\rho_i(t))$ is the purity of $\rho_i(t)$.
%In our case the 2 Renyi entropy is:
\begin{equation}\label{Renyi_axion}
    S_2(t)=-\ln\biggl[1-\frac{\sin^4(2\theta)}{2}\sin^2(\Gamma t)\biggr]
\end{equation}
where $\Gamma=\frac{6A}{r^3}[1+\frac{B}{3}e^{-rm}(3+3rm+r^2m^2]$.
From Eq.~\eqref{Renyi_axion} it is evident that the entanglement is generated by both the dipole-dipole magnetic interaction and by the presence of axion. Neverthless, giving the form of Eq.~\eqref{Renyi_axion}, we can find the time $t=nt^*$ such that
\begin{equation}
\begin{split}
    S_2(nt^*)&=-\ln\biggl[1-\frac{\sin^4(2\theta)}{2}\sin^2\biggl(n\pi(1+\frac{Be^{-mr}}{3}(3+3mr+m^2r^2)\biggr)\biggr]\\
    &\simeq \frac{\sin^4(2\theta)}{2}n^2\pi^2B^2e^{-2mr}(1+mr+\frac{m^2r^2}{3})^2,
    \end{split}
\end{equation}
in which the the dipole-dipole magnetic interaction no longer contributes. Since $B \propto g_p^2$, the Rényi-2 entropy vanishes in the absence of axion interactions ($g_p \rightarrow 0$). Therefore, the observation of entanglement at specific times $t = n t^*$ can be unambiguously ascribed to axion-mediated interactions, thus providing indirect evidence for the existence of axions.

%When entanglement is detected at specific time $t=nt^*$, it can be attributed solely to axion-mediated interactions, providing indirect evidence for the existence of axions.

\section{Conclusion}\label{Conclusions}

The interplay between quantum information theory and high-energy physics can offer novel perspectives and tools to probe fundamental aspects of nature. As paradigmatic example, we have shown that interactions between neutrinos and their surrounding environment can modify the standard neutrino oscillation framework, giving rise to decoherence effects and unconventional oscillation patterns. These deviations from the standard oscillation behavior can, in principle, be detected in neutrino experiments and may provide valuable insight into the underlying properties of neutrinos, such as whether they are Dirac or Majorana particles, and whether new physics mechanisms are at play.
Moreover, we have highlighted that the same decoherence mechanisms may also open a potential pathway to testing the possible violation of CPT symmetry, one of the most fundamental symmetries in quantum field theory and a pillar of the Standard Model. Detecting any indication of CPT violation would have profound implications, requiring a radical revision of our current understanding of fundamental interactions and spacetime structure.

In addition to the study of neutrino dynamics, we have demonstrated how well-known quantities from quantum information science, such as entanglement measures, can be fruitfully employed in the context of high-energy particle interactions. In particular, we have proposed that the generation of entanglement between two fermions, when properly isolated from other interactions, could serve as an indirect probe of axion-mediated forces. This approach complements traditional detection techniques for axions and similar weakly interacting light particles, potentially enhancing sensitivity in parameter regimes that are otherwise challenging to explore with conventional methods.

The ideas presented here underscore the potential of cross-disciplinary approaches that merge concepts from quantum information, particle physics, and quantum field theory. By leveraging the powerful framework of quantum correlations, decoherence, and entanglement dynamics, it becomes possible to design novel experimental strategies and theoretical models that test the limits of the Standard Model and probe possible extensions motivated by unsolved problems such as the nature of dark matter, the matter-antimatter asymmetry, and the possible breakdown of fundamental symmetries. Future investigations should aim to refine these theoretical predictions, quantify experimental feasibility under realistic conditions, and identify specific experimental setups, such as long-baseline neutrino detectors or dedicated entanglement-sensitive measurements, that can maximize sensitivity to these subtle quantum effects. In this way, the synergy between quantum information science and high-energy physics not only deepens our understanding of the quantum structure of fundamental interactions but also opens promising avenues for the discovery of new physics in upcoming experimental programs.

\section*{Acknowledgements}
A.C. and R.S. acknowledge partial financial support from MIUR and INFN. A.C. also acknowledge the COST Action CA1511 Cosmology and Astrophysics Network for Theoretical Advances and Training Actions (CANTATA).

\bibliography{bibliography}

\begin{thebibliography}{10}

\bibitem{Smat}
S.~Weinberg.
\newblock The quantum theory of fields, vol i.
\newblock {\em Cambridge University press}, 2002.

\bibitem{neu1}
Q.~R. Ahmad et~al.
\newblock {Measurement of the rate of $\nu_e+d \to p+p+e^-$ interactions produced by $^8$B solar neutrinos at the Sudbury Neutrino Observatory}.
\newblock {\em Phys. Rev. Lett.}, 87:071301, 2001.

\bibitem{neu2}
Y.~Fukuda et~al.
\newblock {Evidence for oscillation of atmospheric neutrinos}.
\newblock {\em Phys. Rev. Lett.}, 81:1562--1567, 1998.

\bibitem{Pontecorvo}
Samoil~M. Bilenky and B.~Pontecorvo.
\newblock {Lepton Mixing and Neutrino Oscillations}.
\newblock {\em Phys. Rept.}, 41:225--261, 1978.

\bibitem{Pontecorvo1}
S.~M. Bilenky and S.T Petkov.
\newblock Massive neutrinos and neutrino oscillation.
\newblock {\em Review of Modern Physics}, 59(671), 1987.

\bibitem{neu5}
Kanji Fujii, Chikage Habe, and Tetsuo Yabuki.
\newblock {Note on the field theory of neutrino mixing}.
\newblock {\em Phys. Rev. D}, 59:113003, 1999.
\newblock [Erratum: Phys.Rev.D 60, 099903 (1999)].

\bibitem{neu6}
Kanji Fujii, Chikage Habe, and Tetsuo Yabuki.
\newblock {Remarks on flavor neutrino propagators and oscillation formulae}.
\newblock {\em Phys. Rev. D}, 64:013011, 2001.

\bibitem{neu7}
K.~C. Hannabuss and D.~C. Latimer.
\newblock {The quantum field theory of fermion mixing}.
\newblock {\em J. Phys. A}, 33:1369--1373, 2000.

\bibitem{BSM1}
A.~Capolupo, S.~Monda, G.~Pisacane, A.~Quaranta, and R.~Serao.
\newblock {Dark Universe from QFT Mechanisms and Possible Experimental Probes}.
\newblock {\em Universe}, 11(5):142, 2025.

\bibitem{BSM2}
Antonio Capolupo, Simone Monda, Gabriele Pisacane, Raoul Serao, and Aniello Quaranta.
\newblock {Dark matter induced by neutrino mixing and flavor vacuum condensate probed by neutrino capture on tritium}.
\newblock {\em J. Phys. Conf. Ser.}, 3017(1):012041, 2025.

\bibitem{BSM3}
A.~Capolupo, S.~Monda, G.~Pisacane, A.~Quaranta, and R.~Serao.
\newblock {The effect of spacetime torsion on neutrino mixing}.
\newblock {\em J. Phys. Conf. Ser.}, 3017(1):012049, 2025.

\bibitem{BSM4}
Antonio Capolupo, Aniello Quaranta, and Raoul Serao.
\newblock {Exploring X17 and dark charges in the context of Standard Model tensions}.
\newblock In {\em {11th International Workshop on Decoherence, Information, Complexity and Entropy}: {Spacetime - Matter - Quantum Mechanics}}, 3 2025.

\bibitem{BSM5}
Antonio Capolupo, Aniello Quaranta, and Raoul Serao.
\newblock {The impact of the X17 boson on particle physics anomalies: Muon anomalous magnetic moment, Lamb shift and W mass}.
\newblock {\em Phys. Dark Univ.}, 47:101748, 2025.

\bibitem{BSM6}
Antonio Capolupo, Giuseppe De~Maria, Simone Monda, Aniello Quaranta, and Raoul Serao.
\newblock {Quantum Field Theory of Neutrino Mixing in Spacetimes with Torsion}.
\newblock {\em Universe}, 10(4):170, 2024.

\bibitem{BSM7}
Antonio Capolupo, Aniello Quaranta, and Raoul Serao.
\newblock {Field Mixing in Curved Spacetime and Dark Matter}.
\newblock {\em Symmetry}, 15(4):807, 2023.

\bibitem{BSM8}
A.~Capolupo, S.~M. Giampaolo, G.~Lambiase, and A.~Quaranta.
\newblock {Probing quantum field theory particle mixing and dark-matter-like effects with Rydberg atoms}.
\newblock {\em Eur. Phys. J. C}, 80(5):423, 2020.

\bibitem{QG}
Daniele Oriti.
\newblock {\em {Approaches to quantum gravity: Toward a new understanding of space, time and matter}}.
\newblock Cambridge University Press, 3 2009.

\bibitem{QG1}
Claus Kiefer and Barbara Sandhoefer.
\newblock {Quantum cosmology}.
\newblock {\em Z. Naturforsch. A}, 77(6):543--559, 2022.

\bibitem{QG01}
Carlo Rovelli.
\newblock {\em {Quantum gravity}}.
\newblock Cambridge Monographs on Mathematical Physics. Univ. Pr., Cambridge, UK, 2004.

\bibitem{QG02}
Leonardo Modesto and Carlo Rovelli.
\newblock {Particle scattering in loop quantum gravity}.
\newblock {\em Phys. Rev. Lett.}, 95:191301, 2005.

\bibitem{QG2}
Sabine Hossenfelder.
\newblock {\em {Gravity can be neither classical nor quantized}}, pages 219--224.
\newblock 2015.

\bibitem{QG3}
Amjad Ashoorioon, P.~S. Bhupal~Dev, and Anupam Mazumdar.
\newblock {Implications of purely classical gravity for inflationary tensor modes}.
\newblock {\em Mod. Phys. Lett. A}, 29(30):1450163, 2014.

\bibitem{QG4}
Igor Pikovski, Michael~R. Vanner, Markus Aspelmeyer, Myungshik Kim, and Caslav Brukner.
\newblock {Probing Planck-scale physics with quantum optics}.
\newblock {\em Nature Phys.}, 8:393--397, 2012.

\bibitem{QG5}
Andreas Albrecht, Alex Retzker, and Martin~B. Plenio.
\newblock {Testing quantum gravity by nanodiamond interferometry with nitrogen-vacancy centers}.
\newblock {\em Phys. Rev. A}, 90(3):033834, 2014.

\bibitem{Penrose}
Richard Howl, Roger Penrose, and Ivette Fuentes.
\newblock {Exploring the unification of quantum theory and general relativity with a Bose{\textendash}Einstein condensate}.
\newblock {\em New J. Phys.}, 21(4):043047, 2019.

\bibitem{OQS1}
{\textasciiacute}Angel Rivas and Susana~F. Huelga.
\newblock {\em {Open Quantum Systems}}.
\newblock SpringerBriefs in Physics. Springer, 2012.

\bibitem{OQS3}
T.C. Berkelbach and M.~Thoss.
\newblock Special topic on dynamics of open quantum system.
\newblock {\em The journal of Chemistry physics}, 152(020401), 2020.

\bibitem{OQS4}
A.~J. Legget, S.~Chakravarty, A.T. Dorsey, M.P.A. Fisher, A.~Garg, and W.~Zwerger.
\newblock Dynamics of the dissipative two-state system.
\newblock {\em Review of Modern physics}, 59(1), 1987.

\bibitem{Marco1}
Antonio Capolupo, Salvatore~Marco Giampaolo, and Aniello Quaranta.
\newblock {Beyond the MSW effect: Neutrinos in a dense medium}.
\newblock {\em Phys. Lett. B}, 820:136489, 2021.

\bibitem{Marco2}
Antonio Capolupo, Salvatore~Marco Giampaolo, Gaetano Lambiase, and Aniello Quaranta.
\newblock {Discerning the Nature of Neutrinos: Decoherence and Geometric Phases}.
\newblock {\em Universe}, 6(11):207, 2020.

\bibitem{Marco3}
Luca Buoninfante, Antonio Capolupo, Salvatore~M. Giampaolo, and Gaetano Lambiase.
\newblock {Revealing neutrino nature and $CPT$ violation with decoherence effects}.
\newblock {\em Eur. Phys. J. C}, 80(11):1009, 2020.

\bibitem{Marco4}
A.~Capolupo, S.~M. Giampaolo, and G.~Lambiase.
\newblock {Decoherence in neutrino oscillations, neutrino nature and CPT violation}.
\newblock {\em Phys. Lett. B}, 792:298--303, 2019.

\bibitem{Bassi2017}
Angelo Bassi, Andr{\'e} Gro{\ss}ardt, and Hendrik Ulbricht.
\newblock {Gravitational Decoherence}.
\newblock {\em Class. Quant. Grav.}, 34(19):193002, 2017.

\bibitem{Zwiebach2009}
B.~Zwiebach.
\newblock {\em {A first course in string theory}}.
\newblock Cambridge University Press, 7 2006.

\bibitem{Gambini2011}
R.~Gambini and J.~Pullin.
\newblock A first course in loop quantum gravity.
\newblock {\em Cambridge University press}, 2004.

\bibitem{Zurek2007}
B.~Duplantier, J.M. Raimond, V.~Rivasseau, and Editors.
\newblock Quantum decoherence, poincare seminar 2005.
\newblock 2006.

\bibitem{Nielsen2010}
M.~A. Nielsen and I.~L. Chuang.
\newblock Quantum computation and quantum information 10th anniversary edition.
\newblock {\em Cambridge university press}, 2010.

\bibitem{Amico2008}
L.~Amico, R.~Fazio, A.~Osterloh, and V.~Vedral.
\newblock Entanglement in many-body systems.
\newblock {\em Review of Modern physics}, 80(517), 2008.

\bibitem{Solodukhin2011}
S.N. Solodukhin.
\newblock Entanglement entropy of black holes.
\newblock {\em Living Review in relativity}, 14(8), 2011.

\bibitem{Hubeny2015}
V.E. Hubeny.
\newblock The ads/cft correspondence.
\newblock {\em Classical and Quantum gravity}, 32(124010), 2015.

\bibitem{Pei2025}
Junle Pei, Yaquan Fang, Lina Wu, Da~Xu, Mustapha Biyabi, and Tianjun Li.
\newblock {Quantum Entanglement Theory and Its Generic Searches in High Energy Physics}.
\newblock 5 2025.

\bibitem{Buras}
Andrzej~J. Buras.
\newblock {Weak Hamiltonian, CP violation and rare decays}.
\newblock In {\em {Les Houches Summer School in Theoretical Physics, Session 68: Probing the Standard Model of Particle Interactions}}, pages 281--539, 6 1998.

\bibitem{PDG}
S.~Navas et~al.
\newblock {Review of particle physics}.
\newblock {\em Phys. Rev. D}, 110(3):030001, 2024.

\bibitem{Gomez}
J.~J. Gomez-Cadenas, J.~Martin-Albo, M.~Sorel, P.~Ferrario, F.~Monrabal, J.~Munoz-Vidal, P.~Novella, and A.~Poves.
\newblock {Sense and sensitivity of double beta decay experiments}.
\newblock {\em JCAP}, 06:007, 2011.

\bibitem{Dolinski}
Michelle~J. Dolinski, Alan W.~P. Poon, and Werner Rodejohann.
\newblock {Neutrinoless Double-Beta Decay: Status and Prospects}.
\newblock {\em Ann. Rev. Nucl. Part. Sci.}, 69:219--251, 2019.

\bibitem{Elliott}
Steven~R. Elliott and Petr Vogel.
\newblock {Double beta decay}.
\newblock {\em Ann. Rev. Nucl. Part. Sci.}, 52:115--151, 2002.

\bibitem{CPT1}
E.~Lisi, A.~Marrone, and D.~Montanino.
\newblock {Probing possible decoherence effects in atmospheric neutrino oscillations}.
\newblock {\em Phys. Rev. Lett.}, 85:1166--1169, 2000.

\bibitem{CPT2}
R.~Zukanovich~Funchal, E.~M. Santos, W.~J.~C. Teves, and A.~M. Gago.
\newblock {Quantum decoherence and neutrino oscillations}.
\newblock {\em PoS}, HEP2001:216, 2001.

\bibitem{CPT3}
Marcelo~M. Guzzo, Pedro~C. de~Holanda, and Roberto L.~N. Oliveira.
\newblock {Quantum dissipation in a neutrino system propagating in vacuum and in matter}.
\newblock {\em Nucl. Phys. B}, 908:408--422, 2016.

\bibitem{CPT4}
F.~Benatti and R.~Floreanini.
\newblock {Open system approach to neutrino oscillations}.
\newblock {\em JHEP}, 02:032, 2000.

\bibitem{CPT5}
F.~Benatti and R.~Floreanini.
\newblock {Massless neutrino oscillations}.
\newblock {\em Phys. Rev. D}, 64:085015, 2001.

\bibitem{CPT6}
Antonio Capolupo.
\newblock {Total and geometric phases, Majorana and Dirac neutrinos}.
\newblock In {\em {Prospects in Neutrino Physics}}, 4 2019.

\bibitem{AX13}
B.~Ary Dos~Santos Garcia et~al.
\newblock {First mechanical realization of a tunable dielectric haloscope for the MADMAX axion search experiment}.
\newblock {\em JINST}, 19(11):T11002, 2024.

\bibitem{AX14}
Ivan De~Martino, Tom Broadhurst, S.~H. Henry~Tye, Tzihong Chiueh, Hsi-Yu Schive, and Ruth Lazkoz.
\newblock {Recognizing Axionic Dark Matter by Compton and de Broglie Scale Modulation of Pulsar Timing}.
\newblock {\em Phys. Rev. Lett.}, 119(22):221103, 2017.

\bibitem{AX2}
R.~D. Peccei.
\newblock {The Strong CP problem and axions}.
\newblock {\em Lect. Notes Phys.}, 741:3--17, 2008.

\bibitem{AX3}
R.~D. Peccei and Helen~R. Quinn.
\newblock {Constraints Imposed by CP Conservation in the Presence of Instantons}.
\newblock {\em Phys. Rev. D}, 16:1791--1797, 1977.

\bibitem{AX4}
Frank Wilczek.
\newblock {Problem of Strong $P$ and $T$ Invariance in the Presence of Instantons}.
\newblock {\em Phys. Rev. Lett.}, 40:279--282, 1978.

\bibitem{AX5}
S.~B. Treiman and Frank Wilczek.
\newblock {Axion Emission in Decay of Excited Nuclear States}.
\newblock {\em Phys. Lett. B}, 74:381--383, 1978.

\bibitem{AX6}
J.~M. Pendlebury et~al.
\newblock {Revised experimental upper limit on the electric dipole moment of the neutron}.
\newblock {\em Phys. Rev. D}, 92(9):092003, 2015.

\bibitem{AX7}
C.~Abel et~al.
\newblock {Measurement of the Permanent Electric Dipole Moment of the Neutron}.
\newblock {\em Phys. Rev. Lett.}, 124(8):081803, 2020.

\bibitem{AX8}
B.~Graner, Y.~Chen, E.~G. Lindahl, and B.~R. Heckel.
\newblock {Reduced Limit on the Permanent Electric Dipole Moment of Hg199}.
\newblock {\em Phys. Rev. Lett.}, 116(16):161601, 2016.
\newblock [Erratum: Phys.Rev.Lett. 119, 119901 (2017)].

\bibitem{AX9}
V.~V. Flambaum, M.~Pospelov, A.~Ritz, and Y.~V. Stadnik.
\newblock {Sensitivity of EDM experiments in paramagnetic atoms and molecules to hadronic CP violation}.
\newblock {\em Phys. Rev. D}, 102(3):035001, 2020.

\bibitem{AX10}
I.~G. Irastorza et~al.
\newblock {Latest results of the CAST axion search}.
\newblock In {\em {42nd Rencontres de Moriond on Electroweak Interactions and Unified Theories}}, pages 301--310, 3 2007.

\bibitem{AX11}
S.~Andriamonje et~al.
\newblock {An Improved limit on the axion-photon coupling from the CAST experiment}.
\newblock {\em JCAP}, 04:010, 2007.

\bibitem{AX12}
Georg~G. Raffelt.
\newblock {Axions: Motivation, limits and searches}.
\newblock {\em J. Phys. A}, 40:6607--6620, 2007.

\bibitem{AX15}
V.~A. Rubakov.
\newblock {Grand unification and heavy axion}.
\newblock {\em JETP Lett.}, 65:621--624, 1997.

\bibitem{AX16}
Tomohiro Fujita, Kyohei Mukaida, and Tenta Tsuji.
\newblock {Reheating after axion inflation}.
\newblock {\em JCAP}, 07:002, 2025.

\bibitem{AX17}
Michael Dine.
\newblock {Axions: Visible and Invisible}.
\newblock {\em AIP Conf. Proc.}, 93:66--76, 1982.

\bibitem{AX18}
Michael Dine, Willy Fischler, and Mark Srednicki.
\newblock {A Simple Solution to the Strong CP Problem with a Harmless Axion}.
\newblock {\em Phys. Lett. B}, 104:199--202, 1981.

\bibitem{AX19}
Nick Houston, Chuang Li, Tianjun Li, Qiaoli Yang, and Xin Zhang.
\newblock {Natural Explanation for 21 cm Absorption Signals via Axion-Induced Cooling}.
\newblock {\em Phys. Rev. Lett.}, 121(11):111301, 2018.

\bibitem{AX20}
E.~Zavattini et~al.
\newblock {New PVLAS results and limits on magnetically induced optical rotation and ellipticity in vacuum}.
\newblock {\em Phys. Rev. D}, 77:032006, 2008.

\bibitem{AX21}
Pierre Pugnat et~al.
\newblock {First results from the OSQAR photon regeneration experiment: No light shining through a wall}.
\newblock {\em Phys. Rev. D}, 78:092003, 2008.

\bibitem{AX22}
R.~Ballou et~al.
\newblock {New exclusion limits on scalar and pseudoscalar axionlike particles from light shining through a wall}.
\newblock {\em Phys. Rev. D}, 92(9):092002, 2015.

\bibitem{AX23}
Klaus Ehret et~al.
\newblock {New ALPS Results on Hidden-Sector Lightweights}.
\newblock {\em Phys. Lett. B}, 689:149--155, 2010.

\bibitem{AX24}
M.~Arik et~al.
\newblock {Search for Solar Axions by the CERN Axion Solar Telescope with $^3$He Buffer Gas: Closing the Hot Dark Matter Gap}.
\newblock {\em Phys. Rev. Lett.}, 112(9):091302, 2014.

\bibitem{AX25}
S.~Aune et~al.
\newblock {CAST search for sub-eV mass solar axions with 3He buffer gas}.
\newblock {\em Phys. Rev. Lett.}, 107:261302, 2011.

\bibitem{AX26}
A.~Capolupo, G.~Lambiase~and, and G.~Vitiello.
\newblock {Probing mixing of photons and axion-like particles by geometric phase}.
\newblock {\em Adv. High Energy Phys.}, 2015:826051, 2015.

\bibitem{AX27}
A.~Capolupo, I.~De~Martino, G.~Lambiase, and An. Stabile.
\newblock {Axion{\textendash}photon mixing in quantum field theory and vacuum energy}.
\newblock {\em Phys. Lett. B}, 790:427--435, 2019.

\bibitem{AX28}
Antonio Capolupo.
\newblock {Quantum field theory of axion-photon mixing and vacuum polarization}.
\newblock {\em J. Phys. Conf. Ser.}, 1275(1):012052, 2019.

\bibitem{AX29}
R.~Barbieri, C.~Braggio, G.~Carugno, C.~S. Gallo, A.~Lombardi, A.~Ortolan, R.~Pengo, G.~Ruoso, and C.~C. Speake.
\newblock {Searching for galactic axions through magnetized media: the QUAX proposal}.
\newblock {\em Phys. Dark Univ.}, 15:135--141, 2017.

\bibitem{10.1093/acprof:oso/9780199213900.001.0001}
Heinz-Peter Breuer and Francesco Petruccione.
\newblock {\em The Theory of Open Quantum Systems}.
\newblock Oxford University Press, 01 2007.

\bibitem{RevModPhys.81.865}
Ryszard Horodecki, Pawe\l{} Horodecki, Micha\l{} Horodecki, and Karol Horodecki.
\newblock Quantum entanglement.
\newblock {\em Rev. Mod. Phys.}, 81:865--942, Jun 2009.

\bibitem{PhysRev.47.777}
A.~Einstein, B.~Podolsky, and N.~Rosen.
\newblock Can quantum-mechanical description of physical reality be considered complete?
\newblock {\em Phys. Rev.}, 47:777--780, May 1935.

\bibitem{Schrödinger_1935}
E.~Schrödinger.
\newblock Discussion of probability relations between separated systems.
\newblock {\em Mathematical Proceedings of the Cambridge Philosophical Society}, 31(4):555–563, 1935.

\bibitem{PhysicsPhysiqueFizika.1.195}
J.~S. Bell.
\newblock {On the Einstein-Podolsky-Rosen paradox}.
\newblock {\em Physics Physique Fizika}, 1:195--200, 1964.

\bibitem{PhysRevLett.49.1804}
Alain Aspect, Jean Dalibard, and Gerard Roger.
\newblock {Experimental test of Bell's inequalities using time varying analyzers}.
\newblock {\em Phys. Rev. Lett.}, 49:1804--1807, 1982.

\bibitem{PhysRevLett.70.1895}
Charles~H. Bennett, Gilles Brassard, Claude Crepeau, Richard Jozsa, Asher Peres, and William~K. Wootters.
\newblock {Teleporting an unknown quantum state via dual classical and Einstein-Podolsky-Rosen channels}.
\newblock {\em Phys. Rev. Lett.}, 70:1895--1899, 1993.

\bibitem{PhysRevLett.69.2881}
Charles~H. Bennett and Stephen~J. Wiesner.
\newblock {Communication via one- and two-particle operators on Einstein-Podolsky-Rosen states}.
\newblock {\em Phys. Rev. Lett.}, 69:2881--2884, 1992.

\bibitem{BENNETT20147}
Charles~H. Bennett and Gilles Brassard.
\newblock {Quantum cryptography: Public key distribution and coin tossing}.
\newblock {\em Theor. Comput. Sci.}, 560:7--11, 2014.

\bibitem{PhysRevLett.67.661}
Artur~K. Ekert.
\newblock {Quantum cryptography based on Bell's theorem}.
\newblock {\em Phys. Rev. Lett.}, 67:661--663, 1991.

\bibitem{DowlingMilburn2003}
Jonathan~P. Dowling and Gerard~J. Milburn.
\newblock {Quantum Technology: The Second Quantum Revolution}.
\newblock {\em Phil. Trans. Roy. Soc. Lond. A}, 361(1809):1655--1674, 2003.

\bibitem{seesaw}
S.~F. King.
\newblock {Neutrino mass models}.
\newblock {\em Rept. Prog. Phys.}, 67:107--158, 2004.

\bibitem{Simonov2019}
Kyrylo Simonov, Antonio Capolupo, and Salvatore~Marco Giampaolo.
\newblock {Gravity, entanglement and CPT-symmetry violation in particle mixing}.
\newblock {\em Eur. Phys. J. C}, 79(11):902, 2019.

\bibitem{PhysRevLett.83.3081}
Charles~H. Bennett, Peter~W. Shor, John~A. Smolin, and Ashish~V. Thapliyal.
\newblock {Entanglement-assisted classical capacity of noisy quantum channels}.
\newblock {\em Phys. Rev. Lett.}, 83:3081, 1999.

\bibitem{QC1}
Salvatore~M. Giampaolo and Tommaso Macr{\`\i}.
\newblock {Entanglement, holonomic constraints, and the quantization of fundamental interactions}.
\newblock {\em Sci. Rep.}, 9(1):11362, 2019.

\bibitem{QC2}
Charles~H. Bennett, Peter~W. Shor, John~A. Smolin, and Ashish~V. Thapliyal.
\newblock {Entanglement-assisted classical capacity of noisy quantum channels}.
\newblock {\em Phys. Rev. Lett.}, 83:3081, 1999.

\bibitem{QC3}
Antonio Capolupo, Gaetano Lambiase, Aniello Quaranta, and Salvatore~Marco Giampaolo.
\newblock {Probing axion mediated fermion{\textendash}fermion interaction by means of entanglement}.
\newblock {\em Phys. Lett. B}, 804:135407, 2020.

\end{thebibliography}

\end{document}